\documentclass{emulateapj}
\usepackage{apjfonts}

\newcommand{\chandra}{\textit{Chandra}}
\newcommand{\cxo}{\textit{Chandra X-ray Observatory}}

\newcommand{\xmmn}{\textit{XMM-Newton}}

\newcommand{\ergsec}{\ensuremath{\mathrm{erg}~\mathrm{s}^{-1}}}
\newcommand{\Lx}{\ensuremath{L_\mathrm{x}}}

\def\simpropto{\mathrel{\hbox{\rlap{\lower.575ex \hbox {$\sim$}}
                   \kern-.3em \raise.45ex \hbox{$\propto$}}}}

\begin{document}
\shorttitle{\chandra\ observation of GLIMPSE-C01}
\shortauthors{Pooley et al.}
\slugcomment{submitted to ApJL}

\title{{\it Chandra} X-Ray Observation of the Globular Cluster GLIMPSE-C01}

\author{D.\ Pooley\altaffilmark{1}, S.\ Rappaport\altaffilmark{2},  A.\ Levine \altaffilmark{2}, E.\ Pfahl\altaffilmark{3}, J.\ Schwab\altaffilmark{2}}   

\altaffiltext{1}{Astronomy Department, University of Wisconsin--Madison, 475 North Charter Street, Madison, WI  53706, {\tt pooley@astro.wisc.edu }}
\altaffiltext{2}{Department of Physics and Kavli Institute for Astrophysics and Space Research, MIT, Cambridge, MA 02139}
\altaffiltext{3}{Kavli Institute for Theoretical Physics, University of California, Santa Barbara, CA 93106}


\begin{abstract}
We have observed the recently discovered rich star cluster GLIMPSE-C01 for 46 ks with the \cxo. Seventeen X-ray sources with luminosities $\gtrsim 0.6 \times10^{31}$ ergs s$^{-1}$ were discovered, one of which is likely a quiescent low-mass X-ray binary. The spatial distribution of these X-ray sources with respect to the NIR and IR images of the cluster, combined with the luminosity and spectral hardness distributions of the sources, provide strong evidence that GLIMPSE-C01 is a rich Galactic globular cluster.
\end{abstract}



\section{Introduction}

\begin{figure*}[t]
\begin{center}
\includegraphics[width=0.475\textwidth]{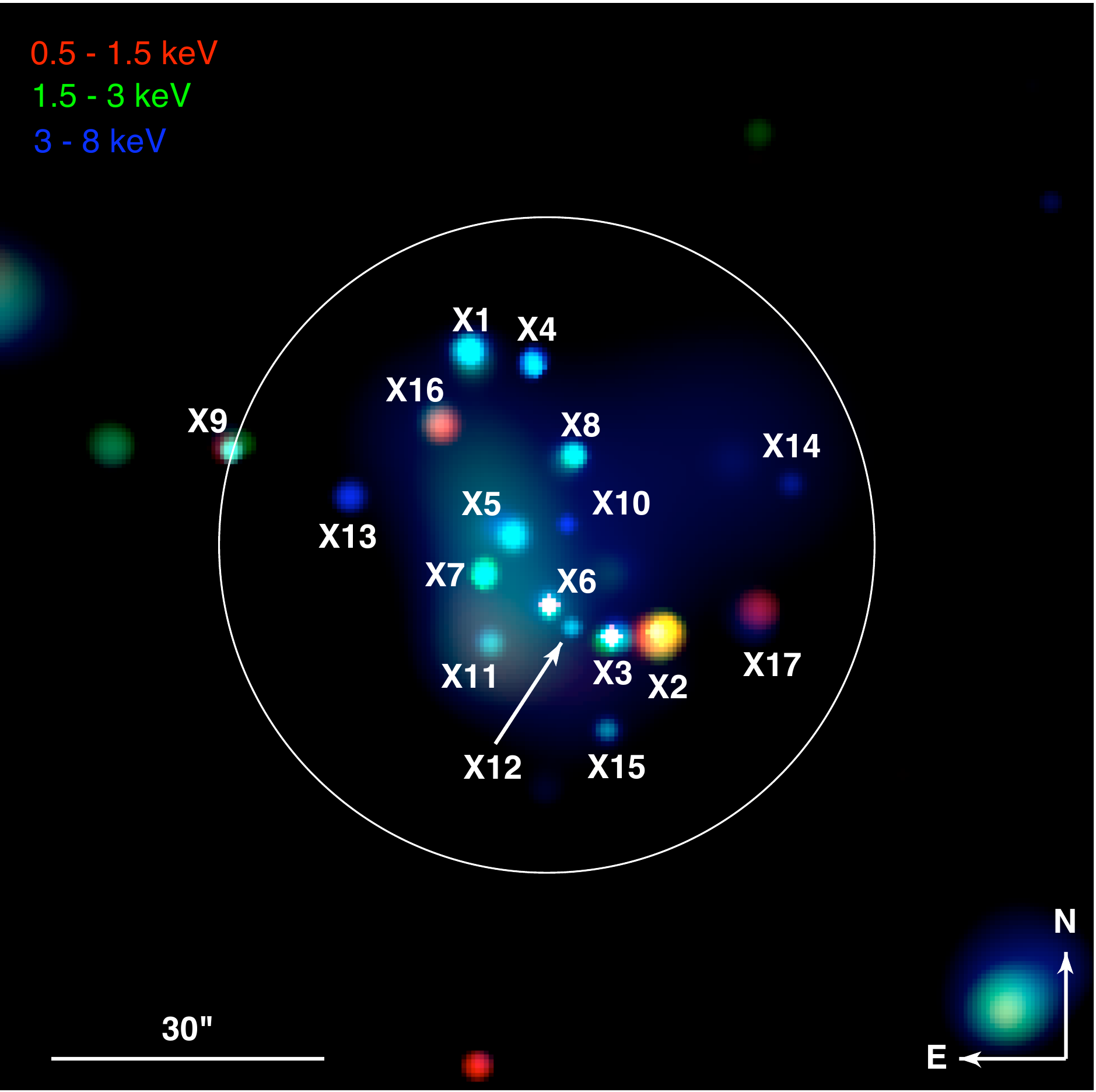}\hglue0.03\textwidth\includegraphics[width=0.475\textwidth]{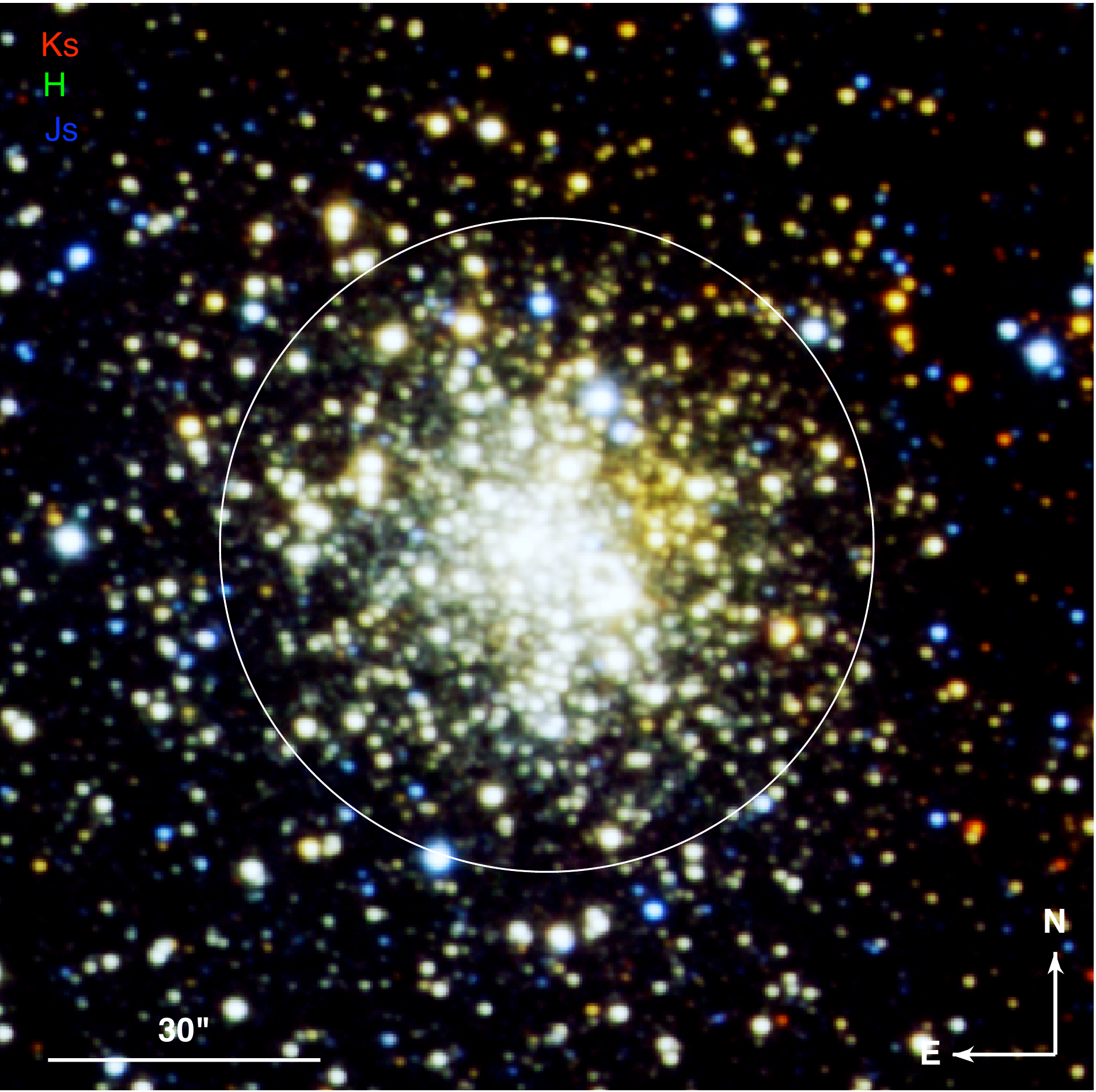}
\caption{$2' \times 2'$ images of the field of GLIMPSE-C01 with the 36\arcsec\ radius half-light region indicated by the white circle.  Left panel: Adaptively smoothed \chandra\ image with sources indicated.  Right panel: True-color image made from the J, H, and K bands of an NTT image (adapted from Ivanov et al.~2005). \label{fig:GC01}}
\end{center}
\end{figure*}

The star cluster ``GLIMPSE-C01'' was discovered only recently with the {\it Spitzer Space Observatory} during the Galactic Legacy Infrared Mid Plane Survey (Kobulnicky et al. 2005), and by cross-correlating {\it ASCA} Galactic-ridge X-ray sources with 2MASS images (Simpson \& Cotera 2004).  This heavily obscured object lies directly in the Galactic plane and is very likely a new rich Galactic globular cluster (see Fig.\,\ref{fig:GC01}).   The cluster center has Galactic coordinates $l = 31^\circ$ and $b = -0.1^\circ$, placing it within 10 pc of the Galactic midplane, as close as any cluster listed in the Harris (1996) catalog\footnote{\url{http://www.physics.mcmaster.ca/Globular.html}}.  The visual extinction is estimated to be $A_V \simeq 12-18$ magnitudes (Kobulnicky et al. 2005), which corresponds to an absorbing column density of $2.7 \pm 0.5 \times 10^{22}$ cm$^{-2}$ (Predehl \& Schmitt 1995).  

Kobulnicky et al.~(2005) found that GLIMPSE-C01 is relatively close to the Earth at a distance of $3-5$ kpc, has a mass of $\sim 1-3 \times 10^5 M_\odot$ and possibly higher, and seems to be highly centrally condensed, with a half-light 
radius of $\sim$0.5--1 pc 
and a core radius that is a factor of $\sim$4 times smaller (see also Ivanov, Kurtev, \& Borissova 2005).  These half-light and core radii are smaller than those for
many
 of the interesting clusters that have previously been studied with {\em Chandra}, implying a very high density in the cluster core and a concomitantly high production rate for X-ray sources.  
 
The possibility that GLIMPSE-C01 might be either an old open cluster or a more distant ``super star cluster'' comprised of many young, massive stars is extensively discussed by Kobulnicky et al.~(2005).  However, the authors rather convincingly discount these alternatives on the basis that (i) no radio emission or
NIR
emission lines are observed, (ii) the cluster has a well-populated giant branch, and (iii) it is located $\approx$ 6 kpc from the Galactic center and old open clusters are extremely scarce at galactocentric radii of $\lesssim 7.5$ kpc (Friel 1995), probably because of the disruptive effects of giant molecular clouds.  GLIMPSE-C01 is the richest cluster uncovered thus far in the {\em Spitzer} midplane survey covering 92 square degrees.  

The field near GLIMPSE-C01 has not previously been observed by \chandra\, but was observed with \xmmn\ for 15 ksec (P.\ Woods, private communication).  In addition, the {\em ASCA} survey of the Galactic Ridge (Sugizaki et al.~2001) lists source AX J184848$-$0219 within $45''$ of the GLIMPSE-C01 cluster center.  
Neither the {\em XMM} nor {\em ASCA} observation had the angular resolution required to resolve the X-ray sources in the core of this cluster.  The {\em XMM} image reveals extended emission at the location of the globular cluster that apparently represents the collection of point sources identified in this work.

Here we report the results of a 46 ks \chandra\ observation of the GLIMPSE-C01 cluster.  Seventeen X-ray sources with luminosities above $\sim$$0.6 \times 10^{31}$ ergs s$^{-1}$ were discovered.  We show that their spatial and spectral properties indicate that GLIMPSE-C01 is a globular cluster as opposed to an open cluster.

\section{{\it Chandra} Observations}

GLIMPSE-C01 was observed for 46 ks on 2006 August 15--16 (ObsID 6587) with the Advanced CCD Imaging Spectrometer (ACIS) on \chandra\ with the telescope aim point on the back-side illuminated S3 chip. The data were taken in timed-exposure mode with the standard integration time of 3.24 s per frame and telemetered to the ground in faint mode.

Data reduction was performed using the CIAO 3.4.1 software provided by the \chandra\ X-ray Center\footnote{\url{http://asc.harvard.edu}}.  The data were reprocessed using the CALDB 3.4.0 set of calibration files (gain maps, quantum efficiency, quantum efficiency uniformity, effective area) including a new 
bad-pixel list made with the acis\_run\_hotpix tool.  The reprocessing was done without including the pixel randomization that is added during standard processing.  This omission slightly improves the point spread function.  The data were filtered using the standard {\it ASCA} grades (0, 2, 3, 4, and 6) and excluding both bad pixels and software-flagged cosmic ray events. Intervals of strong background flaring were searched for, but none were found.  The extraction of counts and spectra and the generation of response files and background spectra were accomplished with ACIS Extract v3.131 (Broos et al.\ 2002), which calls many standard CIAO routines.  All extraction regions were constructed to match the shape of the point-spread function (PSF) and enclose 90\% of the PSF.

\section{X-Ray Image and Source Properties}

The CIAO wavelet-based wavdetect tool was employed for source detection in the 0.5--6.0, 0.3--10.0, and 2--8 keV bands. We detected 42 sources on the entire S3 chip. Of these sources, 12 lie within the 36\arcsec\ half-light radius of GLIMPSE-C01. From the density of sources outside the half-light radius, we expect 0--1 non-cluster sources inside the half-light radius.  We also examined adaptively-smoothed images (made from the CIAO csmooth tool) for possible point sources.  We found five additional candidate sources this way, but some of these may be spurious. The full source list is given in Table 1, with sources numbered according to detected counts in the 0.5--8 keV range.

We fit all sources with absorbed power-law spectral models in Sherpa (Freeman et al.\ 2001) using Cash (1979) statistics, which are appropriate for low count data.  We fixed the column density to $2.7 \times 10^{22}$ cm$^{-2}$.  For each source, we simultaneously fit the source spectrum and associated background spectrum, which was constructed from a source-free annulus around the source.  We performed the simultaneous fitting because Cash statistics are not appropriate for background-subtracted spectra.  The best fit power-law photon index and unabsorbed 0.5--8 keV luminosity, assuming a distance of 3.7 kpc (Ivanov et al.\ 2005) are given in Table 1.

The X-ray image of GLIMPSE-C01 is shown in Fig.\ref{fig:GC01}, along side an NTT image constructed from the J, H, and K bands (courtesy of V. Ivanov \& J. Borissova, private communication; adapted from Ivanov et al.~2005) Both images are color coded for photon energy and intensity.  The X-ray source designations are shown in the \chandra\ image. 

Source positions, luminosities, and spectral properties are summarized in Table 1.  In addition, we list the results of two statistical tests performed by ACIS Extract.  The first is the Kolmogorov-Smirnov probability that the source spectrum and background spectrum are drawn from the same parent distribution; values close to unity indicate they are.  We note that because a background spectrum is constructed from a local annulus around a source and because there is diffuse emission which may be unresolved point sources, a similarity of the source and background spectral shapes is not necessarily sufficient grounds to regard a particular source as spurious.  The second statistic in the table is the Kolmogorov-Smirnov probability that the source's photon arrival times are consistent with a constant-intensity source; values close to zero indicate a variable source.

\begin{deluxetable*}{rllccrcll}
\tablecaption{X-ray Sources in GLIMPSE-C01}
\tablehead{
\colhead{Source} & \colhead{RA (J2000)} & \colhead{Decl.\ (J2000)} & \colhead{Error (\arcsec)\tablenotemark{a}} & \colhead{Counts} & \colhead{\Lx\tablenotemark{b}} & \colhead{PL Index\tablenotemark{c}} & \colhead{$\mathrm{KS}_\mathrm{bkgd}$\tablenotemark{d}} & \colhead{$\mathrm{KS}_\mathrm{constant}$\tablenotemark{e}}}
\startdata
X1 & 18:48:50.261 & $-$01:29:28.57 & 0.07 & 68 &  9.86 &  0.1 & $5.7\!\times\!10^{-8}$ & 0.70 \\
X2 & 18:48:48.867 & $-$01:29:59.56 & 0.07 & 56 &  63.0 & 4.6 & $1.3\!\times\!10^{-12}$ & 0.85 \\
X3 & 18:48:49.231 & $-$01:30:00.05 & 0.08 & 43 &  6.21 &  0.1 & $3.9\!\times\!10^{-3}$ & 0.97 \\
X4 & 18:48:49.799 & $-$01:29:29.96 & 0.09 & 37 &  6.17 &$-$0.4& $3.5\!\times\!10^{-3}$ & 0.64 \\
X5 & 18:48:49.949 & $-$01:29:48.74 & 0.09 & 33 &  4.60 &  0.2 & $1.7\!\times\!10^{-1}$ & 0.14 \\
X6 & 18:48:49.684 & $-$01:29:56.49 & 0.09 & 33 &  4.26 &  0.7 & $2.6\!\times\!10^{-2}$ & 0.71 \\
X7 & 18:48:50.163 & $-$01:29:53.15 & 0.10 & 30 &  4.29 &  1.9 & $9.2\!\times\!10^{-4}$ & 0.06 \\
X8 & 18:48:49.505 & $-$01:29:40.09 & 0.12 & 22 &  3.20 &$-$0.1& $6.9\!\times\!10^{-3}$ & 0.95 \\
X9 & 18:48:52.012 & $-$01:29:39.48 & 0.16 & 14 &  1.73 &  1.2 & $4.9\!\times\!10^{-2}$ & 0.60 \\
X10 & 18:48:49.559 & $-$01:29:47.51 & 0.18 & 10 &  1.22 & 0.5 & $2.6\!\times\!10^{-1}$ & $2\!\times\!10^{-6}$ \\
X11 & 18:48:50.126 & $-$01:30:00.70 & 0.18 & 10 &  1.21 & 1.5 & $8.7\!\times\!10^{-2}$ & 0.47 \\
X12 & 18:48:49.515 & $-$01:29:59.14 & 0.18 & 9 &  1.05 &  1.0 & $2.3\!\times\!10^{-1}$ & 0.29 \\
X13 & 18:48:51.147 & $-$01:29:44.63 & 0.22 & 7 &  1.04 &$-$0.1& $3.7\!\times\!10^{-1}$ & 0.19 \\
X14 & 18:48:47.932 & $-$01:29:43.62 & 0.21 & 7 &  0.51 &    10 & $9.6\!\times\!10^{-1}$ & 0.32 \\
X15 & 18:48:49.243 & $-$01:30:10.21 & 0.22 & 6 &  0.85 &   1.9 & $7.2\!\times\!10^{-1}$ & 0.69 \\
X16 & 18:48:50.505 & $-$01:29:36.45 & 0.28 & 5 &  0.53 &   1.5 & $9.5\!\times\!10^{-1}$ & 0.05 \\
X17 & 18:48:48.183 & $-$01:29:58.35 & 0.25 & 5 &  0.57 &   1.1 & $8.5\!\times\!10^{-1}$ & 0.63 \\
\enddata
\label{fit_param}
\tablenotetext{a}{These statistical uncertainties do not take into account the $\sim$0\farcs4 systematic uncertainty in \chandra's pointing accuracy.}
\tablenotetext{b}{Unabsorbed 0.5--8 keV luminosity in units of $10^{31}$ ergs s$^{-1}$.}
\tablenotetext{c}{Best fit power law photon index.}
\tablenotetext{d}{Kolmogorov-Smirnov probability that the source spectrum and background spectrum are drawn from the same parent distribution; values close to unity indicate they are.}
\tablenotetext{e}{Kolmogorov-Smirnov probability that the source's photon arrival times are consistent with a constant-intensity sources; values close to zero indicate a variable source.}
\end{deluxetable*}

The sources are centrally concentrated around the cluster core, the radius of which we estimate to be $\sim$$9''$ 
(see \S4.2).
This is quite similar to the spatial distribution of low-luminosity X-ray sources found in other Galactic globular clusters (e.g., Pooley et al.~2003; Heinke et al.~2005).  Overall, the range of spectral luminosities, hardness ratios, and spatial distribution of the sources clearly associated with GLIMPSE-C01 are highly reminiscent of a Galactic globular cluster rather than those of an open cluster (see, e.g., Fig.~11 of Kobulnicky et al.~2005; van den Berg et al.~2004).

In addition to the point sources there is also a significant component of diffuse emission nearly aligned with the cluster center and very likely directly associated with the cluster.  
After subtracting out the contribution from the detected point sources and the background, we find an excess of $513\pm45$ counts [0.5--8 keV] within the optical half-light radius. This is comparable to (only about 15\% greater than) the total number of counts from all detected point sources (the sum of the counts listed in Table 1, corrected for the 90\% PSF extraction regions). In the 0.5--2 keV band, the diffuse emission is $82\pm20$ counts, and, in the 2--8 keV band, it is $432\pm40$ counts.
The diffuse X-ray emission has an asymmetric morphology        
that appears to peak near the locus of X7, X5, X6, and X11, which is offset     
about 8\arcsec\ from the optical center of the cluster.  This may be due to     
differential extinction across the field (Kobulnicky et al.\ 2005; Ivanov    
et al.\ 2005); however, even in the 3--8 keV band, which is insensitive to      
extinction, there appears to be asymmetry.  The X-ray excess has a ``half-light'' radius of about 20\arcsec.

\section{Discussion}
\subsection{X-ray Sources}

The 46-ksec \chandra\ observation has allowed us to take a census of the X-ray emitting objects in this cluster (with $L_x \gtrsim 0.6\times10^{31}$ ergs s$^{-1}$).  We find 17 X-ray sources within the half-light radius of the cluster.  From the astrometric comparison that we have made between the X-ray source positions, and the crowded IR and NIR field, we were unable to conclusively associate any of the bright stellar members of the cluster with the X-ray sources; there is one preliminary match, but given the crowded field this 
might 
be a chance superposition.  Since the bright IR sources are likely stellar giants, this lack of a correlation is not unexpected.  Based on identifications of X-ray sources in this luminosity regime in other globular clusters, the counterparts are usually quiescent low-mass X-ray binaries (qLMXBs; Hertz \& Grindlay 1983, Grindlay et al.\ 2001, Pooley et al.\ 2003), cataclysmic variables (CVs; Edmonds et al. 2003, Knigge et al. 2002, Pooley et al.\ 2002), millisecond radio pulsars (MSPs; Freire et al. 2003, Ransom et al.\ 2005), or chromospherically active main-sequence binaries (e.g., BY Draconis systems). 

Above a few times $10^{31}$~\ergsec, we would expect only qLMXBs and CVs, with the qLMXBs spectrally softer than the CVs.  Given the spectral softness of X2 and its intrinsic \Lx, we conclude that it is likely a qLMXB.  The other sources are likely a mixture of CVs, MSPs, and BY Dra systems.  Among these sources, X10 is notable for its extreme variability.  Seven of its 10 detected counts were recorded in a brief 200 s interval, with the other three counts coming in the next 8 ks.  
The luminosity given in Table~1 is the average luminosity throughout the 46 ks observation; during the 200~s flare, X10 achieved $\Lx \approx 2\times10^{33}$~\ergsec.

\subsection{GLIMPSE-C01 Dynamics}
The 100-fold overabundance, per unit mass, of interacting binaries in globular clusters has been known for 30 years (Clark 1975; Katz 1975), and is related to dynamical encounters between collapsed stars (neutron stars, white dwarfs, or black holes) and cluster field stars or binaries (e.g., Hut et al. 1992; Pooley et al. 2003, and references therein).  
The specific types of encounters that have been studied include two-body tidal capture (Fabian, Pringle, \& Rees 1975; Press \& Teukolsky 1977; DiStefano \& Rappaport 1992, 1994), three-body exchange encounters (e.g., Sigurdsson \& Phinney 1995; Rasio, Pfahl, \& Rappaport 2001), and direct collisions (Verbunt 1987; Ivanova et al. 2005).  The total rate, $\Gamma$, of close encounters between two types of objects is given by:
\begin{equation}
\Gamma  \propto \frac{n_{\rm 0,1} n_{\rm 0,2} V}{\sigma} \propto \frac{\rho_0^2 r_c^3}{\sigma} \propto \rho_0^{3/2} r_c^2~~~~,
\end{equation}
where $n_{\rm 0,1}$ and $n_{\rm 0,2}$ are the central densities for the two types of objects, $V$ is the effective volume of the interaction region, and $\sigma$ is the stellar velocity dispersion.  The second proportionality assumes that both $n_{\rm 0,1}$ and $n_{\rm 0,2}$ are proportional to the central mass density, $\rho_0$, and that the relevant interaction volume is that of the cluster core of radius $r_c$; the third proportionality uses the virial relation $\sigma^2 \propto \rho_0 r_c^2$.  

\begin{figure}[t]
\centering
\includegraphics[width=0.47\textwidth]{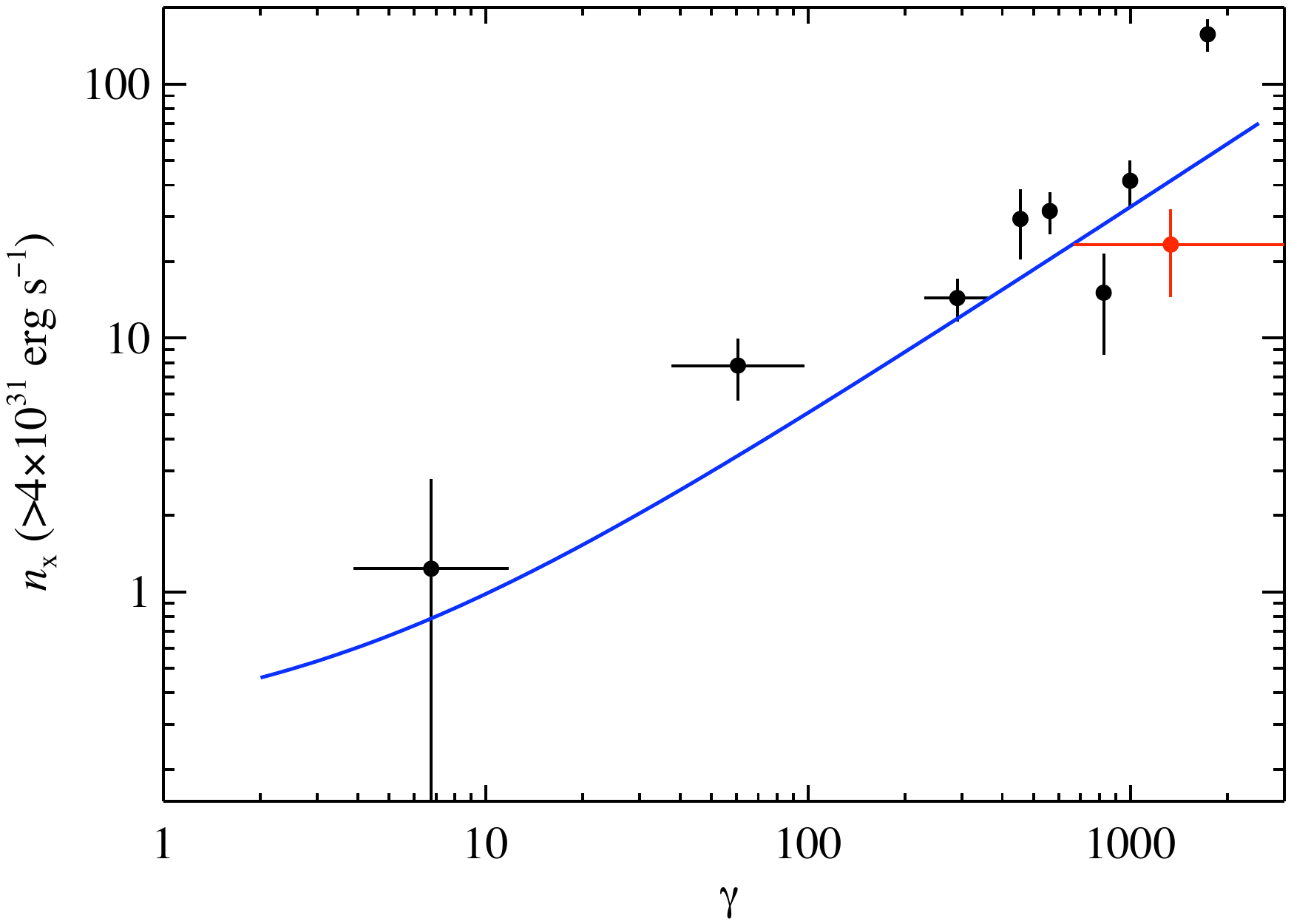}
\caption{Number of globular cluster X-ray sources with $\Lx
\ga 4\times10^{31}$~ergs/sec per unit mass, $n_x$, versus the encounter rate per unit mass
$\gamma$ of the cluster (adapted from Pooley \& Hut 2006).  Two dozen clusters are represented in this figure, but several of the low-$\gamma$ clusters are binned up to improve statistics. The blue line indicates the relationship for the CV-dominated population II of Pooley \& Hut. The red point is a preliminary location on this diagram for the GLIMPSE-C01 cluster, with 7 sources above the \Lx\ limit, $\Gamma = 400^{+600}_{-200}$, and $M=3\times10^5$~M$_\sun$. \label{fig:gamma}}
\end{figure}

Thus far, \chandra\ has been used to observe $\sim$25 Galactic globular clusters, with typical exposure times of $\sim 20-80$ ks; the cluster 47 Tuc has been observed more extensively.  X-ray luminosities at the limits of sensitivity are typically a few $\times 10^{30}$ ergs s$^{-1}$.   A plot of the observed numbers of X-ray sources per unit mass in a given cluster versus the estimated value of $\Gamma$ per unit mass for that cluster is shown in Fig.\,\ref{fig:gamma}.  This shows a strong correlation between the number of X-ray sources in a globular cluster and $\Gamma$.

If we assume that GLIMPSE-C01 has a mass in the range of $M_{\rm cluster} \simeq (1-3) \times 10^5~M_\odot$ we can estimate $\rho_0$ as follows.  From Fig.~10 of Kobulnicky et al.~(2005),
we estimate that the surface brightness falls to $\sim$1/2 its central value at a radial distance of $9''$, and we take this as a good approximation to the core radius,  $r_c$.  We have also constructed a radial light profile from the Ivanov et al.~(2005) NTT image, fit the region out to $\theta = 1'$ with a simple Hubble profile (eq.~2), and find a very similar value for $r_c$ of $\sim$$7'' \pm 2''$.  We note that this value for $r_c$ that we have adopted differs significantly from the values for this parameter cited by Kobulnicky et al.~(2005) and Ivanov et al.~(2005).  
We further take the distance to GLIMPSE-C01 to be $4 \pm 1$ kpc.  This leads to an estimate of the physical size of the core radius of $r_c \simeq 0.17 \pm 0.04$ pc.   For the half-light radius (the radius encircling half the light in the projected image), we adopt the value cited in Table 1 of Kobulnicky et al., viz, $\theta_h \simeq 36''$, or $r_h = 0.70 \pm 0.17$ pc.  Given the roughness of some of these estimates, we feel justified in using the modified Hubble profile (e.g., Binney \& Tremaine 1987) for the cluster surface brightness to relate $\rho_0$ with $r_h$ and $M_{\rm cluster}$:
\begin{equation}
\Sigma(R) = \frac{2 \rho_0 r_c}{[1+(R/r_c)^2]}~~~.
\end{equation}
From this, we find that
\begin{equation}
\rho_0 \simeq \frac{M_{\rm cluster}}{4 \pi r_c^3 \ln[1+(r_h/r_c)^2]}~~~.
\end{equation}
If we plug in the values we adopted above, we find
\begin{equation}
\log \rho_0 = 5.6 \pm 0.3 ~~~[{\rm units ~of}~L_\odot~{\rm pc}^{-3}]~~~,
\end{equation}
where we have assumed a V-band mass to light ratio of 2, and expressed the central density in terms of $L_\odot$ pc$^{-3}$.

We now use these values for $\rho_0$ and $r_c$, and adopt a plausible value of $\sigma = 10$ km s$^{-1}$, to compute a value of the formation rate parameter $\Gamma$ of 
\begin{equation}
\Gamma_{\rm GLIMPSE-C01} = 400^{+600}_{-200}~~~,
\end{equation}
with the same normalization used in Pooley et al.\ (2003), namely that $\Gamma/100$ is roughly the number of LMXBs expected in the cluster.  This value of $\Gamma$, after normalizing to the cluster mass, is plotted in Fig.~2, along with its considerable uncertainties.  As can be seen, GLIMPSE-C01 is consistent with the general trend of other Galactic globular clusters.

Given that we have shown that GLIMPSE-C01 has a high production rate for       
X-ray sources, it appears that this cluster is a good candidate to search       
for msec radio pulsars (Freire et al. 2003; Ransom et al. 2005).

\section{Acknowledgements}
 
D.~P.\ gratefully acknowledges the support provided by NASA through \chandra\ Postdoctoral Fellowship grant PF4-50035 awarded by the \chandra\ X-Ray Center, which is operated by the Smithsonian Astrophysical Observatory for NASA under contract NAS8-03060. S.~R.\ acknowledges support from NASA \chandra\ Grant NAG5-GO6-7072X.  E.~P.\ was supported by NSF Grant PHY05-51164.  J.~S.\ received support from the Paul E.\ Gray (1954) endowed fund for UROP.  The authors thank Fred Baganoff and Mark Bautz for helpful discussions and David Kaplan for initial help in trying to identify optical counterparts. We are grateful to V. Ivanov and J. Borissova for kindly providing us with their NTT image of GLIMPSE-C01.
Finally, we thank Scott Ransom for his participation in the early phases of    
this project.


\end{document}